\documentclass[]{spie}
\usepackage{amssymb}
\usepackage{amsmath}
\usepackage{authblk}
\usepackage{graphicx}
\usepackage{csquotes}
\usepackage{dcolumn}
\usepackage{hyperref}
\usepackage{wrapfig}
\usepackage[utf8]{inputenc}
\usepackage[margin = 0.875in, top = 1.0in, bottom = 1.25in]{geometry}
\usepackage[font=small,labelfont=bf]{caption}

\title{SHIMM as an atmospheric profiler on the Nickel Telescope}
\author[1]{Ollie Jackson}
\author[2]{Maaike A.M. van Kooten}
\author[3]{Saavidra Perera}
\author[1]{Rebecca Jensen-Clem}
\author[1]{Phil Hinz}

\affil[1]{ University of California Santa Cruz}
\affil[2]{National Research Council Canada Herzberg Astronomy and Astrophysics Research Centre}
\affil[3]{University of California San Diego}
\date{}

\begin{document}
\maketitle

\begin{center}
\section*{\large{Abstract}}
\end{center}
\thispagestyle{empty}

Optimal atmospheric conditions are beneficial for detecting exoplanets via high contrast imaging (HCI), as speckles from adaptive optics' (AO's) residuals can make it difficult to identify exoplanets. While AO systems greatly improve our image quality, having access to real-time estimates of atmospheric conditions could also help astronomers use their telescope time more efficiently in the search for exoplanets as well as aid in the data reduction process. The Shack-Hartmann Imaging Motion Monitor (SHIMM) is an atmospheric profiler that utilizes a Shack-Hartmann wavefront sensor to create spot images of a single star in order to reconstruct important atmospheric parameters such as the Fried parameter ($r_0$), $C_n^2$ profile and coherence time. Due to its simplicity, the SHIMM can be directly used on a telescope to get in situ measurements while observing.  We present our implementation of the Nickel-SHIMM design for the one meter Nickel Telescope at Lick Observatory. We utilize an HCIPy simulation of turbulence propagating across a telescope aperture to verify the SHIMM data reduction pipeline as we begin on-sky testing. We also used on-sky data from the AO system on the Shane Telescope to further validate our analysis, finding that both our simulation and data reduction pipeline are consistent with previously determined results for the Fried parameter at the Lick Observatory. Finally, we present first light results from commissioning of the Nickel-SHIMM.

\textbf{Keywords: } astronomical instrumentation, atmospheric profiling, Fried parameter, site testing, High Contrast Imaging
\vspace{10pt}

\section{Introduction}\label{sec:intro}

\hspace{1cm}Over 5000 exoplanets have been detected and confirmed since the first was discovered in the early 1990's, and exoplanet science was one of the most largely discussed fields of research in the 2020 decadal survey \cite{decadal}. From exoplanets, we can learn about the requirements for planet formation, whether the formation of our Solar System is unique, and we can pursue questions of astrobiology and life among the cosmos. There are a number of methods of exoplanet detection that allow astronomers to research these questions, one of which is high contrast imaging (HCI). HCI involves directly taking images of exoplanets through the use of a coronagraph that blocks the light from the host star, allowing for dim objects such as exoplanets to be visible. 

\begin{figure}[h!]
    \centering
    \includegraphics[width = 10cm]{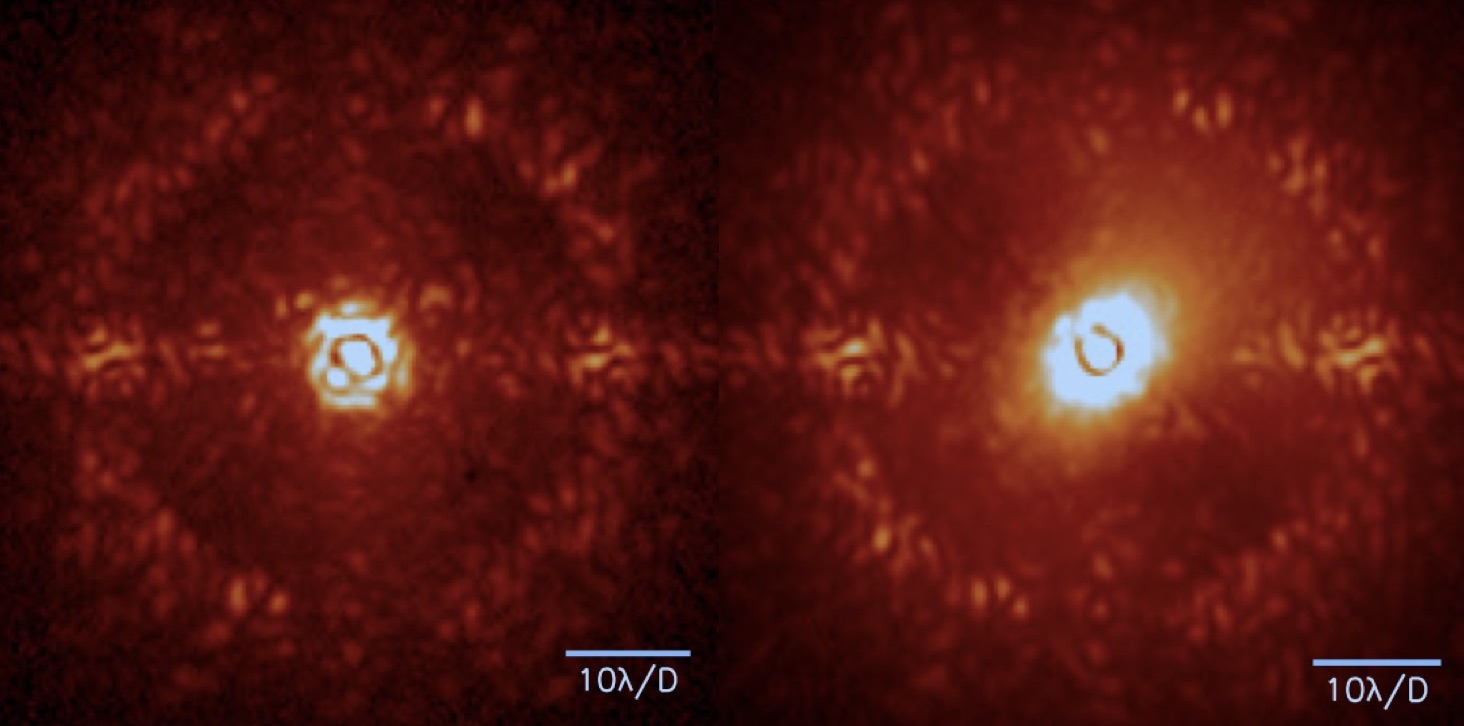}
    \caption{ A night of good atmospheric conditions on the left, and poor atmospheric conditions on the right, with the wind-driven halo effect clear on the right. Taken from Cantalloube \cite{Cantalloube}}
    \label{fig:Cantalloube}
\end{figure}

For ground-based HCI, the instrument's performance is highly dependent on the atmospheric conditions. In Figure~\ref{fig:Cantalloube} from Cantalloube et. al. (2020)\cite{Cantalloube}, the same star is depicted on nights with different conditions, the left from a night with good conditions and the right from a night with poor conditions. On the right, the smearing and spreading of light around the star is referred to as a wind-driven halo, and has been correlated to the atmospheric conditions of high-altitude turbulent layers. The wind-driven halo structure makes it challenging to search the region close to the star for exoplanets, and demonstrates the necessity for optimal atmospheric conditions when using HCI to detect exoplanets with small orbital radii.

Knowing that the atmospheric conditions are sub-optimal just before starting a long exposure on a target would allow astronomers to use their time on science objects that may require less than perfect conditions (for example, searching for an exoplanet further away from a host star on a night of poor conditions, or even focusing on a companion close to a star on a night of excellent conditions). Understanding the on-site atmospheric conditions is then essential for ground-based astronomy because it will allow astronomers to use their telescope time more effectively, and do better science as a result. This has applications beyond just HCI, and can also be applied to queue-based observing such as that done by ESO at the VLT. 

There are a number of seeing monitors and atmospheric profilers that are used frequently in astronomy to determine seeing conditions, namely DIMM (Differential Image Motion Monitor \cite{DIMM}), MASS (Multi Aperture Scintillation Sensor \cite{KornilovMASS}),SCIDAR (SCIntillation Detection And Ranging \cite{SCIDAR}), and SLODAR (SLOpe Detection And Ranging \cite{SLODAR}). DIMM utilizes a mask with two subapertures to get two images of a star. It then calculates the variance of the differential motion between the image locations using the centroids to get an estimate of the Fried parameter ($r_0$) \cite{DIMM}. While DIMM is a seeing monitor that gives us a quantified estimate of the seeing conditions, MASS is an atmospheric profiler that estimates a six-layer atmospheric turbulence profile. By dividing the telescope aperture into rings, MASS can count the number of photons on each ring to determine scintillation patterns and rebuild turbulent strengths at defined heights above the telescope \cite{KornilovMASS}. SCIDAR, like MASS, utilises scintillation information to provide a higher-resolution atmospheric profile by observing. However, unlike DIMM and MASS, it observes two stars as opposed to just one. SCIDAR utilizes the notion that light from two stars with some detectable angular separation must pass through the same turbulence, such that similarly aberrated wavefronts reach the camera from two different stars. A turbulence profile is recreated based on the correlation of how each wavefront is distorted\cite{SCIDAR}. 

SLODAR analyzes where the light from two stars forms images after passing through a Shack-Hartmann Wavefront Sensor (SHWFS) \cite{SLODAR}. Described in further detail in subsequent sections, a SHWFS is comprised of a lenslet array and detector. When light passes through the lenslet array, it focuses into multiple \enquote{spot images}, as visualized in the central row of Figure~\ref{fig:SLODAR v. SHIMM}; an image forms for every lenslet that light passes through. Since SLODAR looks at the light from two stars, it creates two spot arrays (Figure~\ref{fig:SLODAR v. SHIMM}) from the light of each star, and correlations drawn from these spot arrays allow us to rebuild a turbulence profile \cite{Butterley}. 

Each of these previously existing methods of estimating and measuring turbulence profiles and/or measures of the seeing conditions have their benefits and disadvantages. While all provide invaluable information to astronomers, DIMM cannot rebuild a turbulence profile, and both SCIDAR and SLODAR are necessarily large and expensive in order to produce their detailed turbulence profiles. The Shack Hartmann Imaging Motion Monitor (SHIMM) utilizes the same method as SLODAR to acquire $r_0$ and develops on the functionality of DIMM by using a SHWFS instead of a mask to create subapertures \cite{Perera2023}, while simultaneously being a low-cost and portable device that can be more accessibly used by astronomers. The SHIMM's simplicity allows it to be used directly on a telescope to get in situ measurements while observing. 

Here we present our implementation of the SHIMM instrument and optical design onto the 1-meter Nickel telescope (hereafter referred to as Nickel-SHIMM; see Section~\ref{sec:inst design}), as well as the accompanying High Contrast Imaging Python (HCIPy)\cite{por2018hcipy} simulation and SHIMM data reduction pipeline/algorithm in Section~\ref{sec:Algorithm}. We then present the verification of our simulation and data pipeline, followed by instrument commission and first results in Section~\ref{sec:results}.


\section{SHIMM: Shack-Hartmann Imaging Motion Monitor}\label{sec:SHIMM}
\begin{figure}[t]
    \centering
    \includegraphics[width = 15cm]{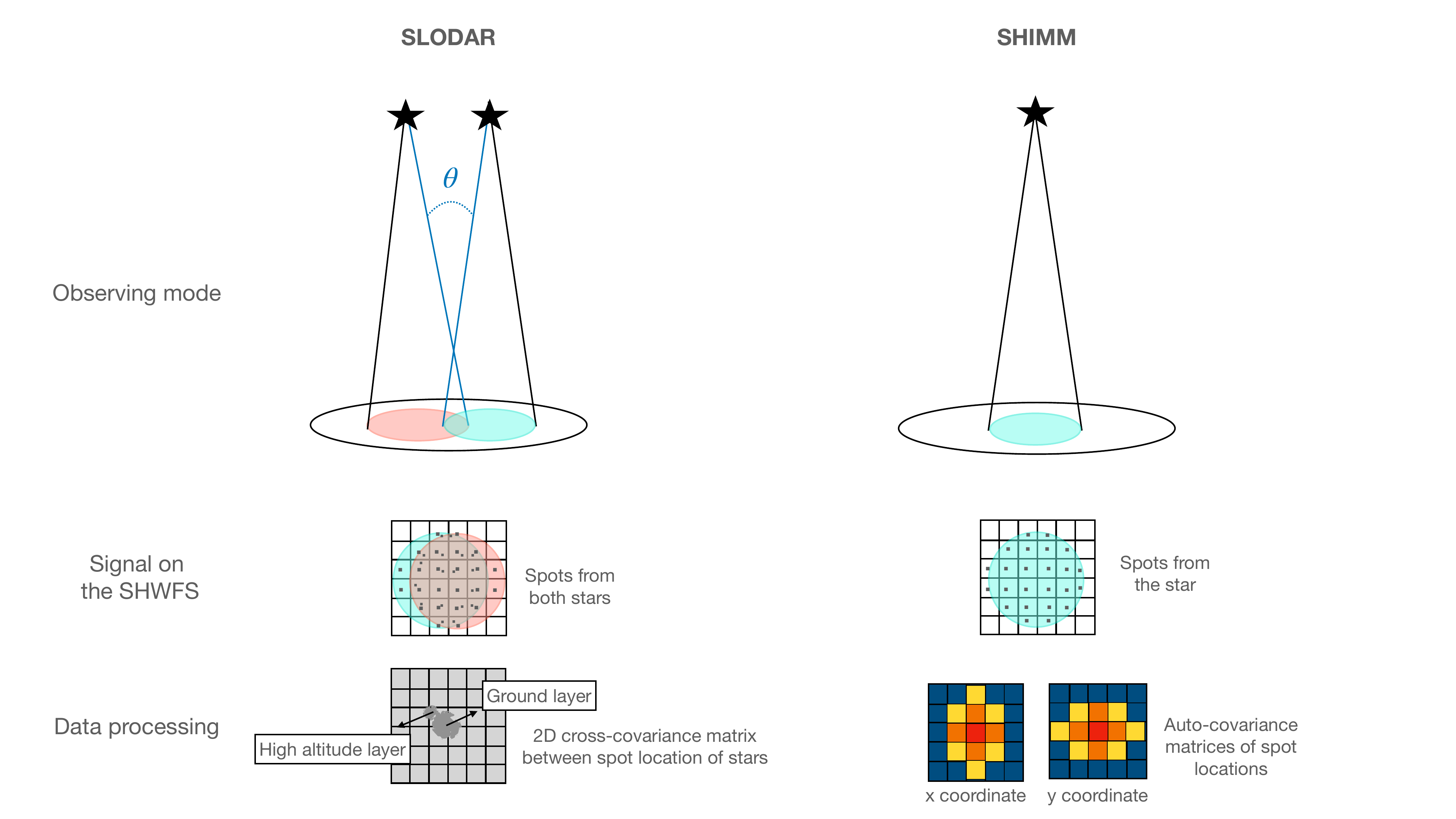}
    \caption{A comparison of the SLODAR and SHIMM designs}
    \label{fig:SLODAR v. SHIMM}
\end{figure}

\hspace{0.5cm} SHIMM is a portable, low-cost instrument that utilizes a single star to calculate and estimate three important parameters that deepen our understanding of atmospheric conditions at the telescope; the Fried parameter ($r_0$), the coherence time ($\tau_0$), and a three-layer profile of atmospheric turbulence as a function of altitude ($C_n^2$ profile) \cite{Perera2023}. The SHIMM utilizes the same methodology as the SLODAR to estimate $r_0$ from subapertures of a single star, while SLODAR and SHIMM use different methodologies to calculate $\tau_0$ and the $C_n^2$ profile. SLODAR uses the correlation of spots from two stars for the $C_n^2$ profile while SHIMM uses only one, as visualized in Figure~\ref{fig:SLODAR v. SHIMM} (noting that SLODAR uses a single star for its estimation of $r_0$). 

Beyond just the physical instrument, SHIMM uses a model of the instrument and a data pipeline that reconstructs the above parameters. In this paper, we focus on SHIMM's calculation of $r_0$ as a measure of atmospheric conditions, and in future research we will describe SHIMM's estimation of $\tau_0$ and $C_n^2$ profile. The $r_0$ value is the length over which the phase of the wavefront changes by 1 radian. This parameter then gives a quantitative description of how turbulent the atmosphere is, and serves as a measure of the atmospheric conditions. 

Since the SHIMM simply utilizes a SHWFS, it can be easily installed onto a telescope (by redirecting some of the science light to the SHIMM) or by using open loop adaptive optics (AO) wavefront sensing data. This provides a unique opportunity to provide atmospheric information directly from the telescope rather than being installed nearby or on an adjacent mountain. As a result, the information we get from SHIMM will directly reflect the behavior of the atmosphere seen by the telescope and gives astronomers an excellent understanding of the on-site atmospheric conditions. Below we outline our implementation of SHIMM on the Nickel telescope (Nickel-SHIMM), which utilizes the same concepts and algorithms as SHIMM to measure $r_0$, $\tau_0$, and the $C_n^2$ profile.

\subsection{The Nickel-SHIMM Instrument Design}\label{sec:inst design}

\begin{figure}[t]
    \centering
    \includegraphics[width = 12cm]{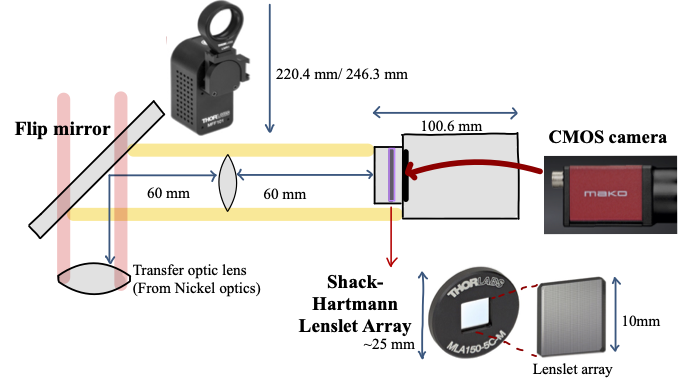}
    \caption{Diagram of the optical design for the Nickel telescope, including visuals of the equipment used. The Nickel's optical pathway is shown in red, and Nickel-SHIMM's in yellow demonstrating functionality of the flip-mirror.}
    \label{fig:Inst. Outline}
\end{figure}

\hspace{0.5cm}The Nickel served as an optimal telescope to demonstrate SHIMM on a larger, observationally driven telescope due to its aperture size and the resulting subaperture size of our lenses. The subaperture size needs to be roughly near the calculated $r_0$ value, and anywhere between 5 cm and 15 cm is a reasonable range for the on-site $r_0$. The Nickel's 40-inch diameter (1.016 m) with 12 subapertures across our lenslet array means that the diameter of each of our subapertures is around 8 cm (1.016 m / 12 subapertures), so it falls well within the 5 cm to 15 cm range of reasonable $r_0$ values. SHIMM's original testing (described in Perera et. al.\cite{Perera2023}) utilized telescope apertures of 9.25 and 11 inches, but the larger the telescope aperture, the more subapertures are available, allowing for a more accurate calculation of $r_0$ with more values to average together\cite{Perera2023}. An even larger telescope than the Nickel would have provided us with more subapertures, but the Nickel also serves as an ideal testing site because it's slightly under-subscribed. The flexibility in the Nickel observing schedule allowed for us to propose engineering time that would have minimal impact on other science being done. 

The Nickel's proximity to the Shane telescope further allows us to compare results from the Nickel to the Shane, and test Nickel-SHIMM on calibration data from Shane's AO system (see Section ~\ref{sec:Shane_test}). This will serve as an excellent check of our results, and is another reason why we proposed to implement SHIMM on the Nickel telescope. In addition, the Lick Observatory is easily accessible from our research institute, allowing for easy travel to the site for installation or in the event of a malfunction.

Our physical implementation of the Nickel-SHIMM instrument on the 1-m Nickel Telescope consists of three main elements; the flip mirror, the Shack-Hartmann lenslet array, and the camera. The flip mirror is an off-the-shelf ThorLabs MFF102, which can be remotely rotated $90^\circ$ in or out of the optical pathway to direct light out of the Nickel's pathway and into our camera. For our instrument, we used the 12.5 mm (0.5 inch) diameter mirror due to limited space within the TUB. The lenslet array (MLA300-14AR from ThorLabs) contains square lenses with a diameter of 295 $\mu$m (pitch of 300 $\mu$m) and focal lengths of 14.3 mm, and is placed in the shaft of our CMOS sensor camera (a Mako G-040 GigE camera from Allied Vision). Figure~\ref{fig:Inst. Outline} displays our equipment and a general diagram of how our instrument is set up. 

The camera and flip mirror are operated via an Intel NUC computer (mounted onto the telescope). To facilitate the power over ethernet required by the camera, we have a smart network switch able to supply power to the camera since it is unsupported by the NUC. This allows us to remotely turn on and off the camera as well, making remote observation possible. 

As mentioned briefly in Section~\ref{sec:intro}, SHIMM utilizes a SHWFS, which consists of a lenslet array and detector (camera). A lenslet array produces multiple images from which we calculate $r_0$. Each lens in the lenslet array samples a small portion of the incoming wavefront and creates its own spot in the focal plane which is called a subaperture. A flat, undistorted (light) wave passing through a lenslet array would form subaperture images on the focal plane at the center of each lenslet, but when the incoming lightwaves are distorted by atmospheric turbulence the images form off-center. The camera (detector) lies in the focal plane of the lenslet array and captures where the subaperture images form. By examining how a subaperture spot moves we can get the local gradient value for that part of the aberrated wavefront. The phase shift of the aberrated wavefront (i.e. how much the incoming lightwave has been distorted from a plane wave) can be reconstructed from these slopes/gradients.

In addition to the flip mirror and SHWFS, our instrument contains a collimating lens (focal length of 60mm) between the flip mirror and lenslet array, which ensures that the lightwaves entering the SHWFS are parallel and reimages the pupil plane for the SHWFS. There is also a transfer optic lens placed before the flip mirror, which is part of the existing Nickel guider optics, and which we then had to account for and incorporate into our design. The transfer optic prescription is undocumented, but we estimate it to have a focal length between 170 and 200 mm based on on-site testing. There is some uncertainty in the focal length of the transfer optic, but it only changes the final location of our pupil plane on the order of a millimeter. We find the best pupil plane during alignment by evenly illuminating the telescope aperture (via the dome flat lamps) to find the sharpest image of the telescope aperture on our camera.


\subsection{SHIMM on the Nickel} \label{sec:Nickel-SHIMM}
\hspace{0.5cm}

\begin{figure}[h!]
    \centering
    \includegraphics[width = 9cm]{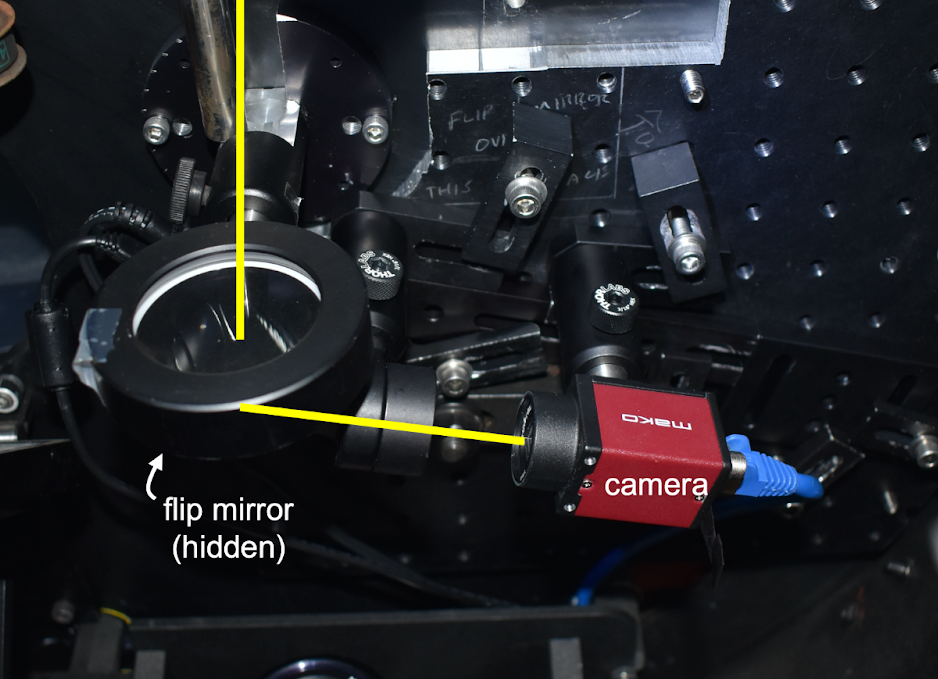}
    \caption{Physical Nickel-SHIMM instrument installed in the TUB of the Nickel telescope. Yellow line indicates Nickel-SHIMM's optical path.}
    \label{fig:shimm on nickel}
\end{figure}

Between the nights of August 14 and August 15, 2022 the instrument was installed onto the 40-inch Nickel Telescope located at the Lick Observatory on Mount Hamilton. Our flip mirror, collimating lens, and camera (containing our Shack-Hartmann lenslet array) were installed in the TUB of the Nickel telescope (which houses the guider arm) at the base of the Cassagrian design, before the Nickel's own CCD camera (Figure ~\ref{fig:shimm on nickel})\footnote{For the Nickel Telescope's hardware overview, see mthamilton.ucolick.org/techdocs/telescopes/Nickel/hw\_overview/}. We adjusted the location of the camera using Vega as a bright star guide to ensure it was centered on the optical path and images from the camera were in focus. Additionally, our NUC computer containing the software necessary to operate the camera and flip mirror was mounted to the outside of the telescope. 

We operate the camera via a Python interface, saving data cubes containing between 500 and 2000 frames and using short exposures of 1-2 ms. The spots images attained are consistent with our simulated predictions, and in a series of frames the effects of turbulence can clearly be seen in the spots (both through movement of the spot centroids and fluctuations in the brightness). Following the procedure outlined in Section~\ref{sec:Algorithm}, the packets of images are easily passed through the SHIMM algorithm to calculate $r_0$.

\begin{center}
\section{\large{HCIPy and SHIMM Algorithm}}\label{sec:Algorithm}
\end{center}

\subsection{HCIPy Simulation}

\begin{figure}[h!]
    \centering
    \includegraphics[width = 15cm]{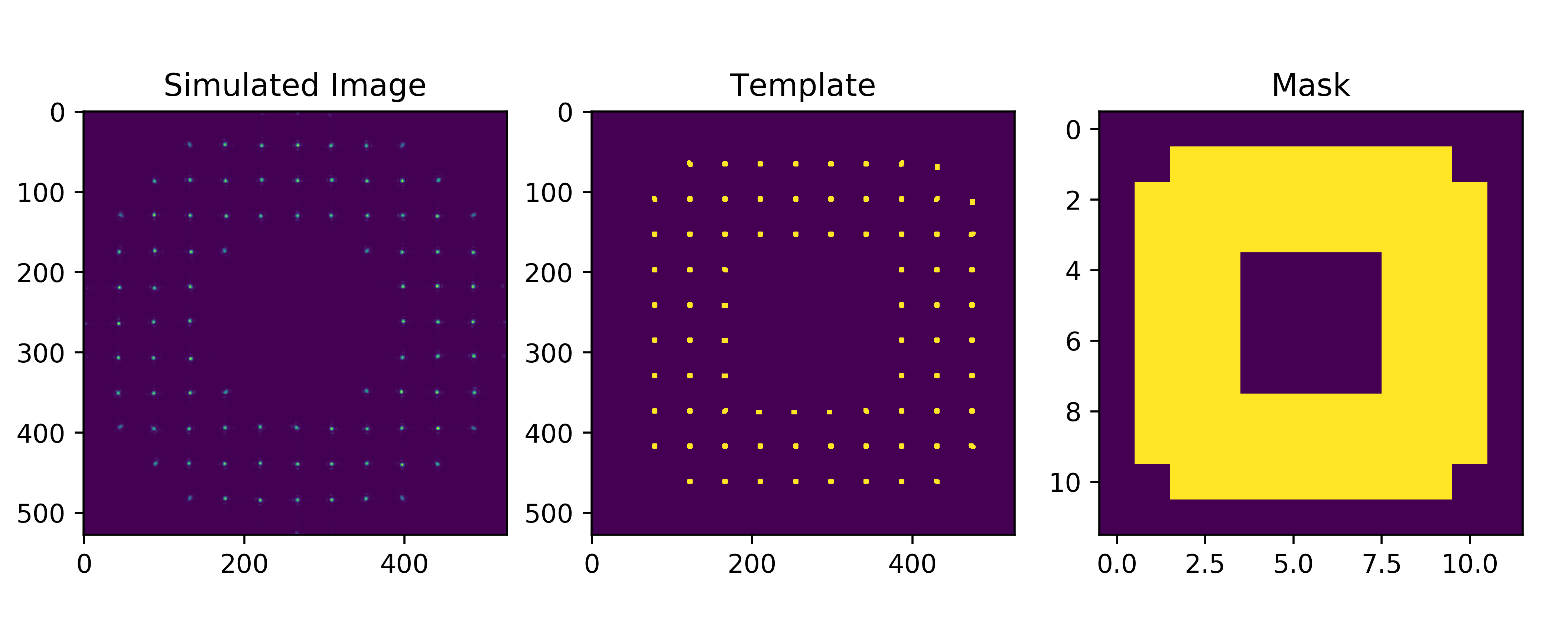}
    \caption{From left to right; a frame from HCIPy simulation, template for spot locations, and mask.}
    \label{fig:template/mask}
\end{figure}

\hspace{0.5cm}We started with an end-to-end simulation of our optical system using the High Contrast Imaging Python (HCIPy) \cite{por2018hcipy} package to simulate light propagating through changing turbulence before passing through the telescope. HCIPy allows us to incorporate the specifications of the telescope (i.e. the diameter of the primary mirror, the central obstruction due to the secondary mirror, the science wavelength of light, and the throughput). For our simulation we assumed a stellar magnitude of 4 and a throughput of 0.5. We additionally assumed an outer scale ($L_0$) of 50 meters and a wind speed of 6 meters per second in both the x and y directions (overall wind speed of roughly 8.5 meters per second). Based on the quantum efficiency of our camera, we chose a wavelength of 500 nm \footnote{See instrument specifications from Allied Vision at www.alliedvision.com/en/camera-selector/detail/mako/g-040/}. We incorporate our SHWFS into the simulation, including the number of lenslets and subaperture size, the focal length of the lenslets, and the lenslet array pitch (see Section~\ref{sec:inst design}). By including the SHWFS in our simulation, we are able to simulate the same subaperture spot images we will get on sky. We define a single frozen-flow atmospheric layer from a set $r_0$ using HCIPy's implementation of an infinite von Karman layer. Finally, we propagate the wavefront through the defined atmospheric layer, and at each timestep capture the image of the wavefront through the simulated SHWFS while evolving the atmospheric turbulence. The result is a stack of simulated spot images, an example of which is shown in the left of Figure~\ref{fig:template/mask}. 

We must ensure that we remove any bias coming from the simulation setup itself. To do this, we use HCIPy to model SHWFS images that haven't been distorted by turbulence, and subtract the resulting slopes off of all simulated images. We start with a singular reference image of our wavefront and a mask, which encodes the spot locations for an undistorted wavefront and maximum spot displacement. We also create a theoretical template of where the images would form from an undistorted wavefront passing through the lenslets. The difference between the mask and the template is that the mask takes each spot where an undistorted image forms and encodes the maximum displacement from that spot based on the diameter of each subaperture (visualized in Figure~\ref{fig:template/mask}). From these (the reference image, the mask, and the template) we locate the spots and centroids of the reference image, and get bias slopes that we must subtract off of our simulated data.

Getting the theoretical/reference slopes is a very similar process to getting the bias slopes; we repeat the process with the wavefront propagating through atmospheric turbulence as opposed to a singular reference wavefront. Our theoretical slopes come from simulated data where we set $r_0$ to be the subaperature diameter of the lenslets (the diameter of the telescope divided by the number of lenslets across our lenslet array). We advance a wavefront through atmospheric layers and take images at each timestep. We do this multiple times with our atmosphere evolving to sufficiently estimate the statics of the system.For every timestep, we get slopes that describe the differential of where the centroid of the image is relative to where it's theoretically supposed to be on our template. These reference slopes, minus the bias slopes, are used to calculate $r_0$ for each real data set. 

\subsection{Extraction of Fried Parameter ($r_0$)}

\begin{figure}
    \centering
    \includegraphics[width = 12cm]{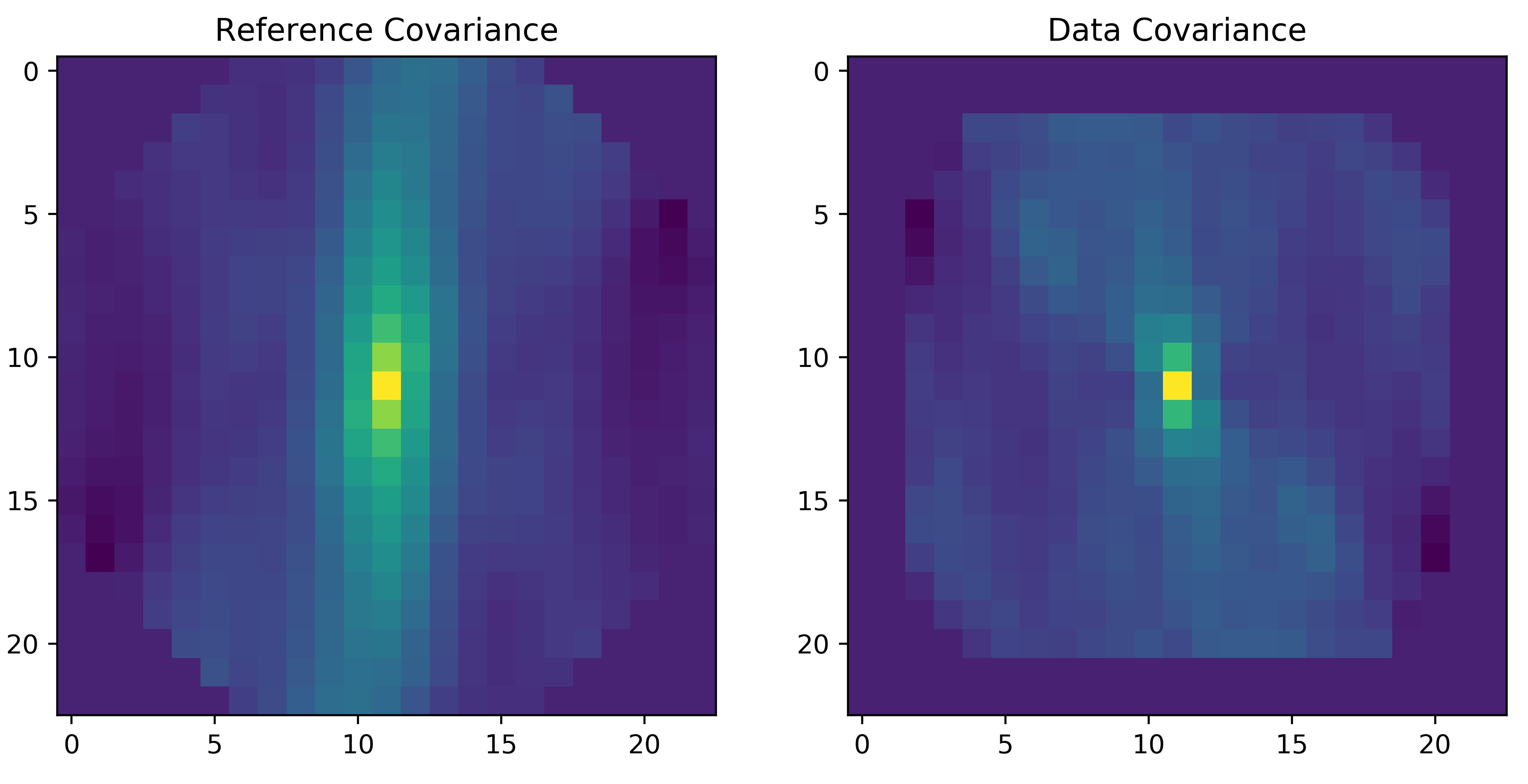}
    \caption{Reference covariance matrix compared to covariance matrix from on-sky data. The central strips are compared to calculate $r_0$.}
    \label{fig:covariances}
\end{figure}

\hspace{1cm} We estimate $r_0$ from the SHWFS subapertures using the methodology outlined in Perera et. al. (2023)\cite{Perera2023} and in Butterley et. al. (2006)\cite{Butterley}. From a time-averaged cube of SHWFS images, we get the measured auto-covariance of the centroids. We additionally create a cube of simulated SHWFS images in the HCIPy simulation, and the auto-covariance of the simulated spot centroids serves as our theoretical model. Theoretical/reference and on-sky auto-covariance maps are created for both the x and y directions. A visual comparison of the auto-covariance maps in the x direction for reference and real data is shown in Figure ~\ref{fig:covariances}. We then fit the theoretical auto-covariance and on-sky auto-covariance maps using a one-dimensional strip through each map as shown in Equation~\ref{eq:r0_i}; we use the central strip, as the signal-to-noise ratio (SNR) decreases as we move away from the center \cite{Perera2023}. A first estimation of $r_0$ for each direction is then found via Equation~\ref{eq:r0_i}, where the $\frac{-3}{5}$ comes from Kolmogorov's power law, $m$ is the fitted parameter (ratio of the measured auto-variance to the theoretical auto-covariance), $s$ is the subaperture diameter of the lenslets in the Shack-Hartmann lenslet array, and $i$ indicates x or y direction. The $r_{0,i}$ values for x and y are then averaged to get $r_0$. 

\begin{equation}
    r_{0i} = m_i^{\frac{-3}{5}}*s
    \label{eq:r0_i}
\end{equation}

We must also take into consideration the airmass above the telescope in order to get $r_0$. We do this using the Pyephem library, which (given the date and time of our observation, location of the telescope, and star name) provides the zenith angle of our star. In our analysis, we noted that Pyephem's date and time was an hour behind Universal Standard time, and had to adjust our calculation of the zenith angle as a result. The airmass is given by 1/cosine(zenith angle), and the final $r_0$ value considering airmass is given in Equation ~\ref{eq:r0_air}, where $r_0$ in Equation~\ref{eq:r0_air} is the average of the two $r_{0,i}$ values calculated via Equation~\ref{eq:r0_i}.

\begin{equation}
    r_{0,\text{air}} = r_0 * \text{airmass}^{\frac{-3}{5}} 
    \label{eq:r0_air}
\end{equation}

\subsection{Verifying Nickel-SHIMM and HCIPy Simulation}
\hspace{0.5cm} In order to verify the accuracy of our simulation, we created multiple simulated datasets using various $r_0$ values. The difference between these simulated datasets and our the simulated data we used to get our reference slopes is determined entirely by $r_0$; an $r_0$ value equal to the subaperture diameter of the lenslets defines our theoretical values. We created simulated data sets with $r_0$ values ranging from 7 cm to 25 cm, then for each data we recalculated $r_0$ from the spot images as described in the previous section.

\begin{figure}[h!]
    \centering
    \includegraphics[width = 10cm]{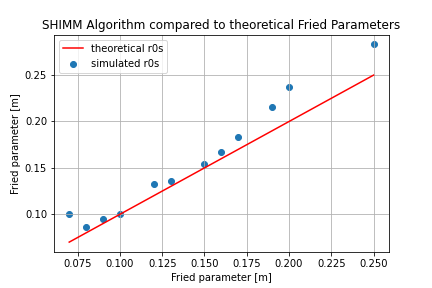}
    \caption{Reconstructed r0 values from SHIMM algorithm (blue dots) compared with the "real" input r0 values (in red)}
    \label{fig:r0 test}
\end{figure}

Figure ~\ref{fig:r0 test} shows the theoretical $r_0$ values used to create the data plotted against the $r_0$ values recalculated using the SHIMM algorithm. Our simulation produces realistic results up to around 18 cm, and then it begins to stray from where we would expect. For our purposes, however, an $r_0$ value within one to two centimeters of the actual value is sufficient to accurately estimate the atmospheric conditions, and therefore we consider our simulation to be working correctly. 

\subsection{Testing on Shane AO Data}\label{sec:Shane_test}
\hspace{1cm} We also had access to calibration data for the AO system on the Shane telescope also located on Mount Hamilton. The Shane telescope has an AO system that allows it to correct for the effects of atmospheric turbulence in science images. When the Shane's AO system is run in \enquote{open loop} (the deformable mirror is not correcting for turbulence), its SHWFS is measuring the full atmospheric turbulence. A short amount of this open loop data is taken for calibration purposes when Shane's AO is used for observing, and occasionally this data gets saved, allowing us to test our SHIMM algorithm and calculate $r_0$. We adjusted our HCIPy simulation and SHIMM algorithm to fit the specifications of the Shane telescope, which also involved creating new reference images, templates, and masks. 

\begin{figure}[h!]
    \centering
    \includegraphics[width = 12cm]{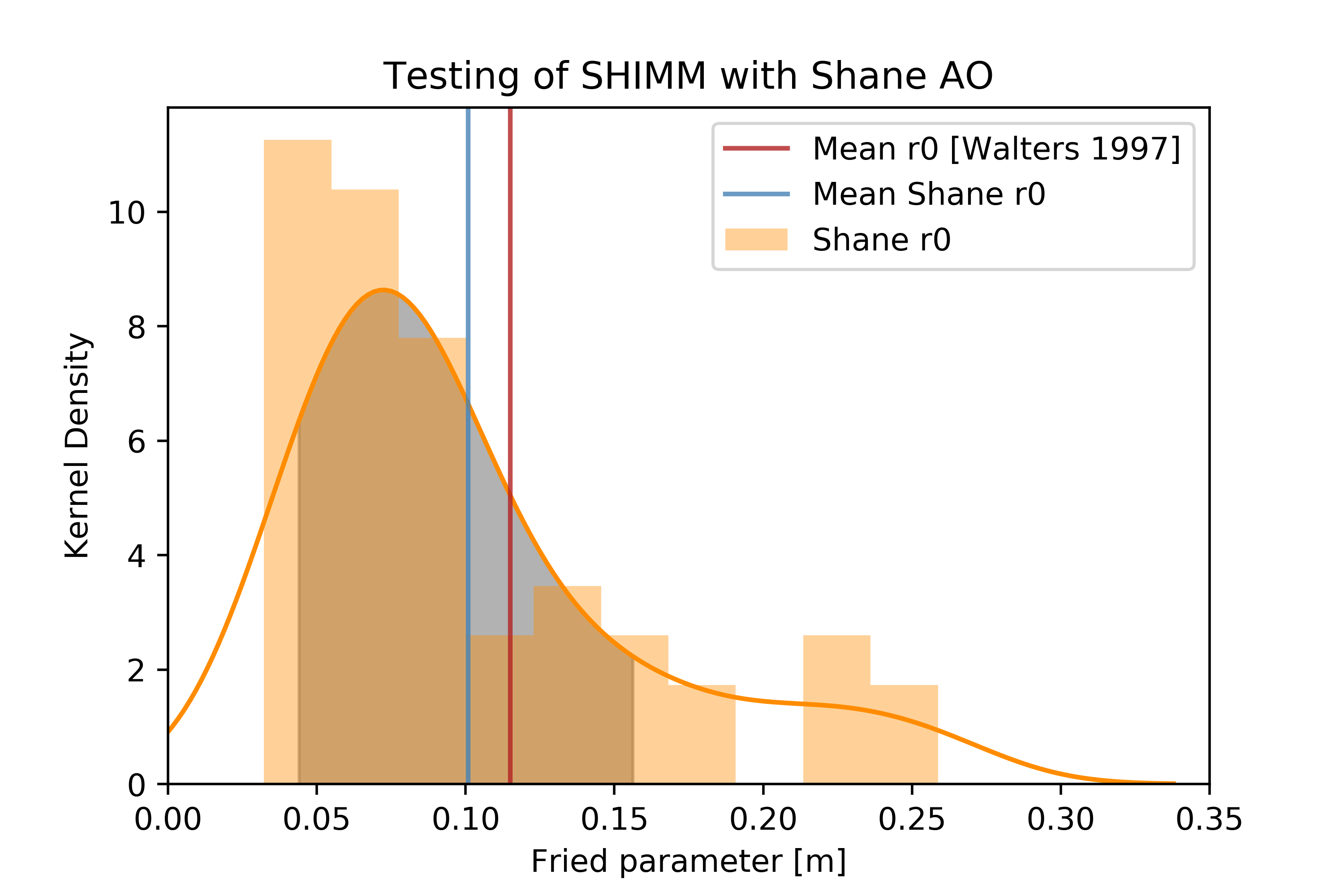}
    \caption{Histogram of calculated $r_0$ values overlaid with Kernel density function (estimation of the probability density using a non-parametric model). In red is the mean $r_0$ determined by Walters\cite{Walters}, which is consistent to a $\sigma$ range (in grey) with the mean $r_0$ on the Shane (shown in blue). }
    \label{fig:shane test}
\end{figure}

Using approximately 15 nights of data we created a histogram of the calculated $r_0$ values, and compared this with the average $r_0$ on Mount Hamilton as determined previously in Walters (1997)\cite{Walters}. The calculated mean from our nights of data using our SHIMM algorithm was 10.1 cm $\pm$ 5.7 cm, and the mean determined in Walters (1997)\cite{Walters} was 11.5 centimeters \cite{Walters}. The value from Walters falls within a $\sigma$ standard deviation range around our mean; our calculated values are consistent with other determinations of $r_0$ at the Lick Observatory (see Figure~\ref{fig:shane test}). 

The SHIMM algorithm's ability to recalculate $r_0$ from simulated data to a reasonable degree, and verify the average $r_0$ value for the Lick Observatory using on-sky data, reinforces that both the simulation and algorithm are working as expected and are ready for on-sky testing. 

\vspace{0.5cm}
\begin{center}
  \section{\large{On-Sky Testing and Results}}  \label{sec:results}
\end{center}

\hspace{0.5cm} We were granted frequent observation time on the Nickel from August 2022 through July 2023, typically collecting about an hour of data each night just before sunrise. The results presented here are from data collected between May and July of 2023. We focused our analysis on the star Deneb, and the binary system Alberio was used to calibrate our system and determine the pixel scale (which we found to be roughly 2 arcseconds per pixel). The exposure and gain of our camera was adjusted nightly, but data was consistently collected using exposure times between 1-2 milliseconds.

\begin{figure}[h!]
    \centering
    \includegraphics[width = 12cm]{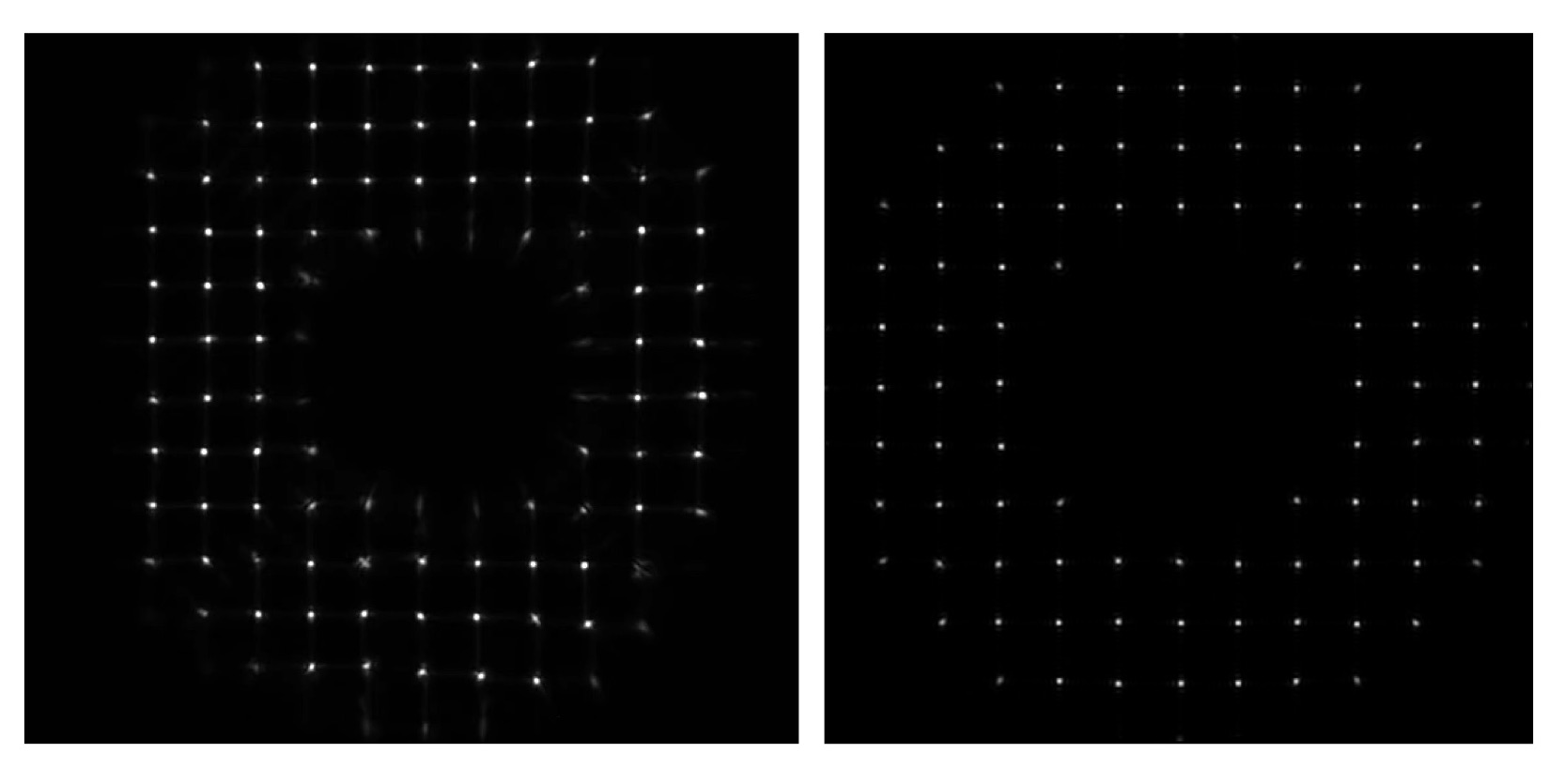}
    \caption{Left: Image captured by our camera during first light. Right: Image from HCIPy simulation of the Nickel.}
    \label{fig:first light}
\end{figure}

\begin{figure}
    \centering
    \includegraphics[width = 12cm]{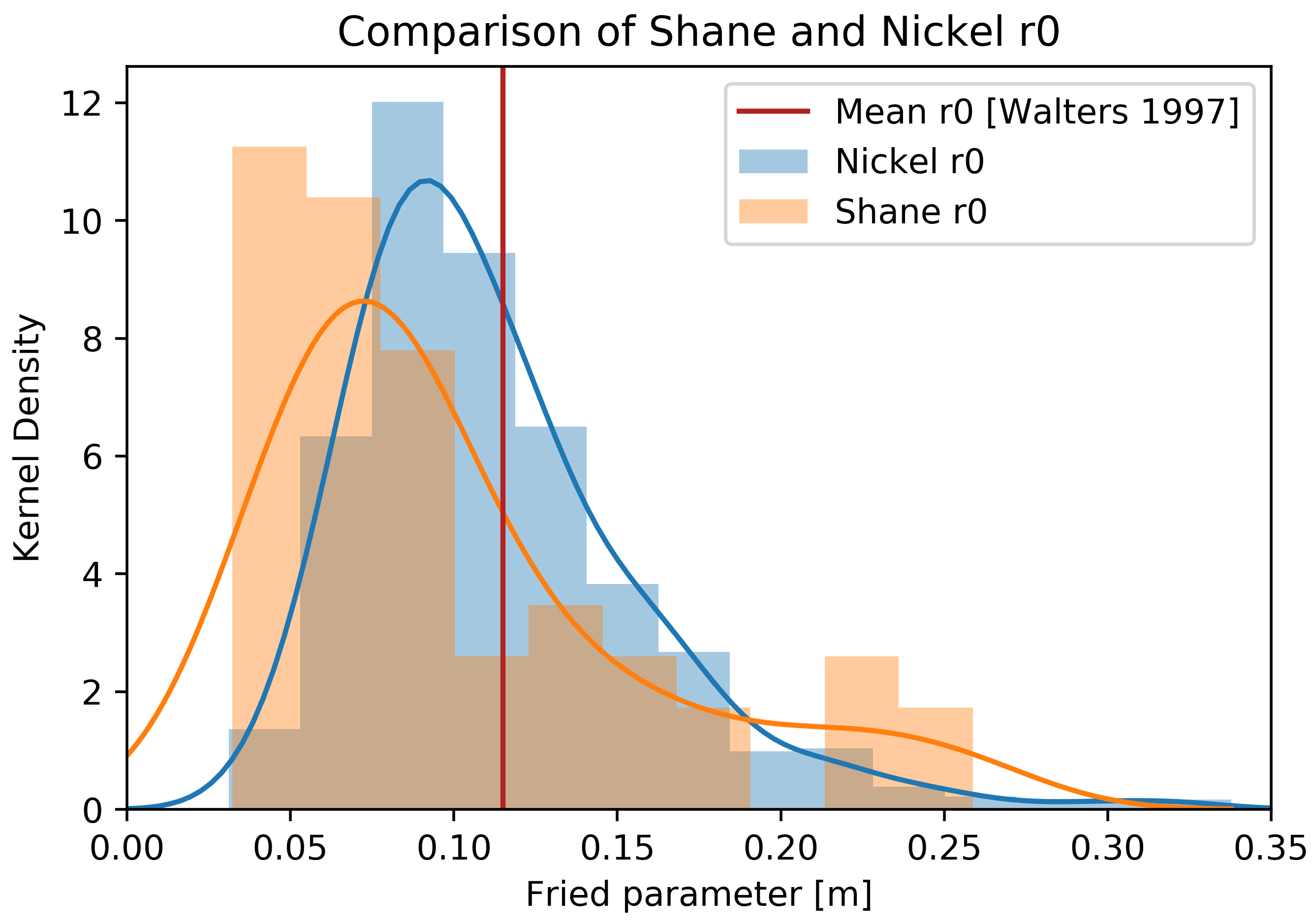}
    \caption{Preliminary results of Nickel-SHIMM compared with tests of SHIMM algorithm on the Shane telescope. Histograms for each group of results overlaid with corresponding Kernel Density functions, with the mean $r_0$ value calculated in Walters (1997)\cite{Walters} again shown in red.}
    \label{fig:results}
\end{figure}

We calculated $r_0$ on data cubes of 500 frames at a time, using the procedure outlined above. Preliminary results are displayed alongside our $r_0$ estimations for the Shane telescope in Figure ~\ref{fig:results}. Across our nights of data, we found the average $r_0$ value to be 11.6 cm $\pm$ 5.6 cm, which is consistent with the mean $r_0$ of 11.5 cm determined by Walters (1997)\cite{Walters}, and is further consistent with the results from testing Nickel-SHIMM on Shane AO calibration data. 

In Figure~\ref{fig:r0 test} we saw that above $r_0$ values of 18 cm that our SHIMM pipeline began to overestimate the actual $r_0$ value. We stated that for our purposes, this behavior would not drastically skew our results. This is corroborated by the fact that our mean $r_0$ values are well below 18 cm, and that the conditions at the Nickel rarely reach $r_0$ values above 18 cm.

\section{Discussion and Conclusions}

\hspace{1cm} Our results for $r_0$ at the Nickel are statistically consistent with both our tests of the SHIMM algorithm on Shane open-loop adaptive optics data, and with the mean $r_0$ for the Lick Observatory in Walters (1997) \cite{Walters}. This gives us confidence that our implementation of Nickel-SHIMM is working correctly as we continue testing and optimizing our analysis. While the preliminary results for the Shane telescope and Nickel telescope are consistent, the $r_0$ values calculated from Shane data display different behavior than those from the Nickel telescope. There are a handful of reasons for this, first that we had much more on-sky data for the Nickel telescope than for the Shane telescope. It's also important to note that our testing of the SHIMM algorithm on Shane open-loop data only provided a test for our data pipeline and simulation; we did not have timestamps or the star target for the Shane data so we were unable to take the airmass above the telescope into account. Finally, with the Shane data, we have the deformable mirror for which the quality of the flatness might affect our measurements. 

As visualized in Figure ~\ref{fig:first light}, our camera and lens are slightly off-axis, resulting in a tilt in our images compared to our simulation. When on-sky images were compared with the templates, the spots in the upper and left sides consistently overlapped well with their theoretical locations, but the spots in the bottom right side were on average further from their theoretical locations. This likely results in an under-estimation of $r_0$, where atmospheric conditions are determined to be worse than they really are, because a number of subapertures are always abnormally far from their theoretical, undistorted locations. In future analysis, these subapertures will be cut when calculating $r_0$. 

As mentioned, Nickel-SHIMM is able to provide a three-layer $C_n^2$ profile and $\tau_0$, in addition to $r_0$. These capabilities are currently in development for our implementation of Nickel-SHIMM. These results indicate that Nickel-SHIMM is operating as expected and is able to provide accurate representations of the atmospheric conditions using in-situ measurements. As expected, Nickel-SHIMM will be able to provide real-time information of atmospheric conditions to astronomers in future implementations. 

\section{\large{Acknowledgements}}
\hspace{1cm} This research was made possible in part from the National Science Foundation (award no. 1852393) and the Lamat Fellowship for Undergraduate Research at the University of California Santa Cruz (UCSC). We would also like to thank Philip Rice for his key support of the on-site implementation of SHIMM. This work was done in collaboration with the Center for Adaptive Optics and the Lab for Adaptive Optics at UCSC. Thank you to Rebecca Jensen-Clem and Eleanor Gates, who provided us with the open-loop data from the Shane AO system. Thank you to UCO and the UCO Machine Shops at UCSC, particularly Pavl Zachary and all the staff at Mt Hamilton who operate the telescopes there. 

The University of California Santa Cruz is located on the unceded land of the Awaswas speaking Uypi tribe, and is currently stewarded under the hands of the Amah Mutsun Tribal Band. Mount Hamilton is located on the lands of the Muwekma, Ohlone, and Tamien/Tamyen Nations. This research would not be possible without the lands on which it was completed, and those lands belong to their historic and current Indigenous peoples.

\bibliography{report} 
\bibliographystyle{spiebib} 

\end{document}